\documentclass[twocolumn,showpacs,pra,amsmath,amssymb]{revtex4}

\usepackage{graphicx}
\usepackage{dcolumn}
\usepackage{bm}
\usepackage{multirow}

\begin{document}

\preprint{APS/123-QED}

\title{Exact Solution of Bogoliubov Equations for Bosons in One-Dimensional Piecewise Constant Potential}

\author{Daisuke Takahashi}%
 \email{takahashi@vortex.c.u-tokyo.ac.jp}
\affiliation{%
Department of Basic Science, University of Tokyo, Tokyo 153-8902
}%

\date{\today}

\begin{abstract}
	We show that Bogoliubov equations in one-dimensional systems with piecewise constant potentials can be always solved. In particular, we analyze in detail the case where the condensate wavefunction is a real-valued function, and give the explicit expressions for wavefunctions of Bogoliubov excitations. By means of these solutions, we consider transmission and reflection properties of Bogoliubov excitations for two types of potential, namely, a rectangular barrier and a potential step. The results yield simple and exact examples of anomalous tunneling effect and quantum evaporation.
\end{abstract}

\maketitle

\section{Introduction}
	\begin{figure}[bt]
		\includegraphics{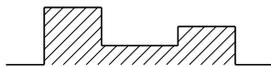}
		\caption{\label{fig1} An example of piecewise constant potential.}
	\end{figure}
	If the first step toward understanding of a given physical system is to elucidate the ground state properties, the next step would be the elucidation of the low-energy excited states. Indeed, experimentally accessible response to an external field is calculated from the information of excited states and linear response theory. In many condensed matter systems, low-energy excited states consist of not only single particle excitations but also collective excitations. As is well known that collective mode in a Bose-Einstein condensate(BEC) is well described by Bogoliubov theory\cite{bogoliubov}.\\
	\indent Bose-Einstein condensation in ultracold atomic gases has been stimulating many theorists ever since its experimental realization\cite{Anderson,Davis,Bradley}. Bogoliubov theory has been also well confirmed in these systems by the observation of sound propagation\cite{Andrews} and the measurement of static structure factor\cite{Stamper,Steinhauer}. Another fascinating issue in ultracold atomic gases is the realization of low-dimensional systems. In one-dimension, in particular, a variety of physical regimes, from BEC regime to Tonks-Girardeau(TG) gas regime, are realized\cite{Gorlitz,Greiner,Moritz,Paredes,Kinoshita}. Theoretically, BEC regime can be attacked by mean-field theory and is often referred to as ``quasi-one-dimensional''\cite{Carr,CarrClark}, while TG regime must be treated as a strongly correlated system. (Exceptionally, modified mean field theory for TG regime has been proposed in Ref.~\cite{Kolomeisky}, and its validity has been investigated \cite{GirardeauWright}.) In this sense, the analysis of Bogoliubov excitations in a one-dimensional system, which we will show in the present paper, belongs to the studies of quasi-one-dimensional systems.\\
	\indent In the theoretical works of quasi-one-dimensional systems, not only the dynamics of the Bose condensate by using Gross-Pitaevskii(GP) equation\cite{Gross,Pitaevskii}, but also the physical properties of Bogoliubov excitations have been investigated by using Bogoliubov equations. For example, perfect transmission in the low-energy limit (known as ``anomalous tunneling'') and related or extended tunneling problems\cite{kovrizhinmaksimov,kovrizhin,kagan,danshitayokoshikurihara,KatoNishiwakiFujita,TsuchiyaOhashi,WatabeKato,OhashiTsuchiya,TsuchiyaOhashi2,takahashikato}, excitation spectrum and dynamical instability in an optical lattice\cite{danshitakuriharatsuchiya,danshitatsuchiya}, and Anderson localization of Bogoliubov excitations in a random potential\cite{BilasPavloff}. Recently, considering the junction of BECs with different interaction strengths, an analog of Andreev reflection\cite{ZapataSols} and Hawking radiation\cite{RecatiPavloffCarusotto} are discussed.\\
	\indent Though the Bogoliubov equations are solved in many works numerically or analytically, few exact solutions are known. Probably, the solution under a dark soliton or a gray soliton\cite{chen,kovrizhin} is the most important one, in the sense that not a few works\cite{danshitayokoshikurihara,danshitakuriharatsuchiya,danshitatsuchiya,OhashiTsuchiya} have used this solution to solve the scattering problem against the potential barrier with the shape of a delta-function. There is no doubt that the delta-function barrier models have succeeded in clarifying the fundamental physics of Bogoliubov excitations. However, several works\cite{KatoNishiwakiFujita,WatabeKato} suggest that transmission and reflection properties of low-energy Bogoliubov excitations are universal, irrespective of the shape of potential. It is therefore highly desirable that such universal properties are supported not only by numerical evidence or zero-width potential (i.e., delta function) but also by exactly solved models in the presence of barrier with finite width and potential steps. In addition, the set of exact solutions will be useful when one considers a new physical problem.\\
	\indent In this paper, we exactly solve Bogoliubov equations in a one-dimensional system in the presence of piecewise constant potential shown in Fig. \ref{fig1}. With use of these exact solutions, we study the transmission and reflection properties of Bogoliubov excitations in two particular examples. Stationary GP equation with this kind of potential shape has been already solved generally\cite{CarrClark,seaman}, but Bogoliubov equations have not been solved.\\ 
	\indent This paper is organized as follows. In Sec.~\ref{secreal}, we show the solution of Bogoliubov equations for the real-valued condensate function. In Sec.~\ref{secapply}, we treat two particular examples, i.e., a rectangular barrier and a potential step, and solve the transmission and reflection problems. In Sec.~\ref{sec:comp}, we extend the solution when the condensate wavefunction is complex-valued. In Sec.~\ref{secout}, we discuss our results and future problems. The conclusion is given in Sec.~\ref{seccon}.

\section{Solutions for the Real-Valued Condensate Wavefunction}\label{secreal}
\subsection{Fundamental Equations}
	We begin with the following one-dimensional time-dependent Gross-Pitaevskii(GP) equation:
	\begin{align}
	\begin{split}
	\mathrm{i}\hbar\frac{\partial}{\partial t}\psi(x,t)&=\left(\!-\frac{\hbar^2}{2m}\frac{\partial^2}{\partial x^2}\!+\!U(x)\!\right)\!\psi(x,t)\\
	&\qquad\qquad\qquad\qquad+g|\psi(x,t)|^2\psi(x,t).
	\end{split}
	\end{align}
	In this paper, we consider only the repulsive interaction, i.e., $ g>0 $. This equation can be rewritten in a dimensionless form by introducing the following quantities:
	\begin{align}
	\begin{split}
		&\bar{x} = \frac{x}{\xi},\quad \bar{t} = \frac{\hbar}{m\xi^2}t,\quad \bar{U}(\bar{x})=\frac{m\xi^2}{\hbar^2}U(x), \\
		&\text{and }\ \bar{\psi}(\bar{x},\bar{t})=\frac{\sqrt{mg}\xi}{\hbar}\psi(x,t),
	\end{split}\label{eq:xi}
	\end{align}
	where $ \xi $ is an arbitrary positive constant with dimension of length, and often taken to be a healing length. Henceforth, we mainly use these dimensionless quantities, and omit bars. Time-dependent GP equation then becomes
	\begin{align}
	\mathrm{i}\frac{\partial}{\partial t}\psi(x,t)=\left(\!-\frac{1}{2}\frac{\partial^2}{\partial x^2}\!+\!U(x)\!\right)\!\psi(x,t)+|\psi(x,t)|^2\psi(x,t).
	\end{align}
	Setting the condensate wavefunction in the form of
	\begin{equation}
	\psi(x,t)={\rm e}^{-\mathrm{i} \mu t}\!\left\{\Psi(x)+\left[u(x)\mathrm{e}^{-\mathrm{i} \epsilon t}-v^*(x)\,\mathrm{e}^{\mathrm{i} \epsilon t}\right]\right\} \label{eq: smallpsi}
	\end{equation}
	and taking the terms up to first order with respect to $u(x), v(x)$, we obtain the stationary GP equation
	\begin{gather}
		\hat{L}\Psi(x)=0,\ \hat{L} = -\frac{1}{2}\frac{\mathrm{d}^2 }{\mathrm{d} x^2} +U(x)-\mu+ |\Psi(x)|^2 \label{eqGP}
	\end{gather}
	for the condensate wavefunction and Bogoliubov equations 
	\begin{gather}
		\begin{pmatrix}\! \hat{L}+|\Psi(x)|^2 & -(\Psi(x))^2 \! \\ \! -(\Psi(x)^*)^2 & \hat{L}+|\Psi(x)|^2 \! \end{pmatrix}\!\! \begin{pmatrix}\! u(x)\! \\ \!v(x)\! \end{pmatrix} = \epsilon \!\begin{pmatrix}\! u(x)\! \\ \!-v(x) \! \end{pmatrix}\! \label{eqBogo}
	\end{gather}
	for the wavefunctions of excitations. Needless to say, these equations can also be derived by diagonalization of mean field Hamiltonian.(e.g., \cite{fetter,griffin}) \\
	\indent In this section, we consider only the case where the condensate wavefunction is a real-valued function for simplicity. Discussion on a complex-valued condensate will be given in Sec.~\ref{sec:comp}. Taking account of $ \Psi(x) $ being real, GP equation can be simplified as
	\begin{align}
		\hat{H}\Psi(x) = 0,\ \hat{H} = -\frac{1}{2}\frac{\mathrm{d}^2 }{\mathrm{d} x^2}+U(x)-\mu+\Psi(x)^2. \label{eq:GP}
	\end{align}
	Further, introducing
	\begin{align}
		S=u+v,\quad G=u-v,
	\end{align}
	Bogoliubov equations are rewritten as
	\begin{align}
		\hat{H}S &= \epsilon G, \label{eq:bogoS} \\
		(\hat{H}+2\Psi^2)G &= \epsilon S. \label{eq:bogoG}
	\end{align}
	$ S $ and $ G $ can be interpreted as phase and density fluctuations, respectively, as pointed out in Ref.~\cite{fetterrokhsar}.
\subsection{Main Results}
	From now, we consider the solution of Bogoliubov equations for a piecewise constant potential. It suffices for our purpose to consider general solutions for only one interval with a constant potential, since the solution valid for all regions can be obtained by joining solutions of each region smoothly. Thus, we set $ U(x)=U_0=\text{const} $. In this situation, GP equation (\ref{eq:GP}) can be immediately integrated once:
	\begin{align}
		(\Psi')^2 = \Psi^4+2(U_0-\mu)\Psi^2+C_{\text{GP}}, \label{eq:GPonce}
	\end{align}
	where the constant of integration $ C_{\text{GP}} $ is determined from boundary conditions. This equation can be integrated once again, and the solution generally becomes Jacobi elliptic function.\\
	\indent By using $ C_{\text{GP}} $ in Eq. (\ref{eq:GPonce}), we can summarize the solution of Bogoliubov equations as follows:\\[0.5em]
	\textit{\indent There exists the particular solution such that $ G $ is proportional to $ S' $, and $ S $ satisfies the following first-order differential equation:}
	\begin{gather}
		G = \frac{1}{\mathrm{i}K}\frac{\mathrm{d}S}{\mathrm{d}x}, \label{eq:gds} \\
		\Bigl(\Psi^2\!+\!\frac{\epsilon^2}{K^2}\Bigr)\frac{\mathrm{d}S}{\mathrm{d}x}-\biggl[ \frac{\mathrm{i}\epsilon}{K}\Bigl(\frac{K^2}{2}\!+\!U_0\!-\!\mu\!+\!\Psi^2\Bigr)+\Psi\frac{\mathrm{d}\Psi}{\mathrm{d}x} \biggr]S = 0. \label{eq:1storderS}
	\end{gather}
	\textit{Here the constant $ K $ satisfies the following quartic equation:}
	\begin{align}
		\epsilon^2 = \left( \frac{1}{2}K^2+U_0-\mu \right)^2-C_{\text{GP}}. \label{eq:detk}
	\end{align}
	\\
	\indent Equation (\ref{eq:detk}) determines four possible values of $ K $, which is consistent with the number of linearly independent solutions for the original system of equations (\ref{eq:bogoS}) and (\ref{eq:bogoG}). The solution of Eq.~(\ref{eq:1storderS}) can be always written down explicitly even when $ \Psi $ is an elliptic function, by using incomplete elliptic integral of the third kind. See Appendix \ref{appjacobi} and the example for a rectangular barrier in Sec. \ref{secapply}.\\ 
	\indent We note that this method is applicable even when energy $ \epsilon $ is negative or complex, so Landau instability and dynamical instability can be also discussed.\\
	\indent In the rest of this subsection, we prove the results summarized above. Eliminating $ G \text{ or } S $ from Eqs. (\ref{eq:bogoS}) and (\ref{eq:bogoG}), we obtain the following fourth-order differential equations for $ S \text{ or } G $: 
	\begin{align}
		(\hat{H}+2\Psi^2)\hat{H}S&=\epsilon^2 S, \label{eq:bogoS2}\\
		\hat{H}(\hat{H}+2\Psi^2)G&=\epsilon^2 G. \label{eq:bogoG2}
	\end{align}
	We can show that if $ U(x)=U_0=\text{const} $, the following operator identity holds:
	\begin{align}
		\frac{\mathrm{d}}{\mathrm{d}x}(\hat{H}+2\Psi^2)\hat{H}=\hat{H}(\hat{H}+2\Psi^2)\frac{\mathrm{d}}{\mathrm{d}x}, \label{eq:opeiden}
	\end{align}
	which can be proved by a straightforward calculation with use of Eq.~(\ref{eq:GP}). From this identity and Eq.~(\ref{eq:bogoS2}), we obtain
	\begin{align}
		\hat{H}(\hat{H}+2\Psi^2)\frac{\mathrm{d}S}{\mathrm{d}x} = \epsilon^2 \frac{\mathrm{d}S}{\mathrm{d}x},
	\end{align}
	which means that $ S' $ is the solution of (\ref{eq:bogoG2}), that is, equation for $ G $. Therefore, there exists the particular solution which satisfies the relation (\ref{eq:gds}). Substituting (\ref{eq:gds}) to Eqs. (\ref{eq:bogoS}) and (\ref{eq:bogoG}), one obtains
	\begin{align}
		\hat{H}S &= \frac{\epsilon}{\mathrm{i}K}S', \label{eq:bogoS3} \\
		(\hat{H}+2\Psi^2)S' &= \mathrm{i}K \epsilon S. \label{eq:bogoG3}
	\end{align}
	Eliminating $ S''' $ and $ S'' $ from (\ref{eq:bogoS3}), (\ref{eq:bogoG3}), and derivative of (\ref{eq:bogoS3}), one obtains Eq.~(\ref{eq:1storderS}). Furthermore, using (\ref{eq:GPonce}), (\ref{eq:1storderS}), (\ref{eq:bogoS3}), and derivative of (\ref{eq:1storderS}), one can obtain Eq.~(\ref{eq:detk}).
\subsection{Special Cases: Solutions Expressed in terms of Elementary Functions}
	When $ C_{\text{GP}}=0 \text{ or } (U_0-\mu)^2 $,  $ \Psi(x) $ becomes an elementary function, and the solutions of Bogoliubov equations also become elementary functions. In these cases, the constant of proportionality between $ G $ and $ S' $ should be taken in a slightly different form rather than (\ref{eq:gds}) to write down the solutions neatly.\\
	\indent When $ C_{\text{GP}}=(U_0-\mu)^2 $, $ \Psi' = \pm(\Psi^2+U_0-\mu) $ follows from Eq. (\ref{eq:GPonce}), and
	\begin{align}
		S&=\mathrm{e}^{\mathrm{i}kx}\biggl( \Psi\pm\frac{\mathrm{i}k}{2} \biggr), \\
		G&=-\frac{\mathrm{i}k}{2\epsilon}S'
	\end{align}
	are solutions of Bogoliubov equations. Here $ k $ obeys the following equation:
	\begin{align}
		\epsilon^2 = \frac{1}{4}k^2\bigr(k^2-4(U_0-\mu)\bigl).
	\end{align}
	When $ C_{\text{GP}}=0 $, on the other hand, the solutions are given by
	\begin{align}
		S&=\mathrm{e}^{\mathrm{i}kx}\biggl( \frac{\Psi'}{\Psi}+\frac{\mathrm{i}\bigl(k^2-2(U_0-\mu)\bigr)}{2k} \biggr), \\
		G&=-\frac{\mathrm{i}\bigl(k^2+2(U_0-\mu)\bigr)}{2\epsilon k}S',
	\end{align}
	and $ k $ satisfies
	\begin{align}
		\epsilon^2 = \biggl( \frac{1}{2}k^2+U_0-\mu \biggr)^2.
	\end{align}
	All possible elementary solutions are summarized in Table \ref{table:elem}. 
	\begin{table*}[htbp]
		\begin{center}
		\caption{\label{table:elem} List of solutions expressed in terms of elementary functions. They are classified by the sign of $ U_0-\mu $ and the value of $ C_{\text{GP}} $.  $ a,\,b $ in the table represent a non-zero real value. If one replaces $ x\rightarrow x-\mathrm{i}\pi/(2a) $ or $ a\rightarrow \mathrm{i}a $ in (ii), one obtains (i) or (iv), respectively. Likewise, $ x\rightarrow x-\pi/(2a) $ in (iv) gives (iii), $ x\rightarrow x-\pi/(2b) $ in (v) gives (vi), and $ b\rightarrow\mathrm{i}b $ in (vi) gives (vii). The last (viii) is the limiting case ($ a,b\rightarrow0 $) of (ii) and (vii).}
		{\small
		\begin{tabular}{|c|c|c|c|c|c|c|}
		\hline
		& $ C_{\text{GP}} $ & $U_0\!-\!\mu$ & $ \Psi(x) $ & $ S(x) $ & $ G(x) $ & Equation for $ k $ \\
		\hline\hline
		(i) & \multirow{4}*[-2em]{$ a^4 $} & \multirow{2}*[-0.9em]{$-a^2<0$} & $\displaystyle a\tanh(ax) $ & $\displaystyle \vphantom{\Bigg|} \mathrm{e}^{\mathrm{i}kx}\left( a\tanh(ax)-\frac{\mathrm{i}k}{2} \right) $ & \multirow{4}*[-2.0em]{$\displaystyle -\frac{\mathrm{i}k}{2\epsilon}S' $} & \multirow{2}*[-0.9em]{$\displaystyle \epsilon^2 = \frac{1}{4}k^2\left( k^2\!+\!4a^2 \right) $} \\
		\cline{1-1}\cline{4-5}
		(ii) & &  & $\displaystyle a\coth(ax) $ & $\displaystyle \vphantom{\Bigg|} \mathrm{e}^{\mathrm{i}kx}\left( a\coth(ax)-\frac{\mathrm{i}k}{2} \right) $ &  & \\
		\cline{1-1}\cline{3-5}\cline{7-7}
		(iii) & & \multirow{2}*[-0.9em]{$a^2>0$} & $ a\tan(ax) $ & $\displaystyle \vphantom{\Bigg|} \mathrm{e}^{\mathrm{i}kx}\left( a\tan(ax)+\frac{\mathrm{i}k}{2} \right) $ & & \multirow{2}*[-0.9em]{$\displaystyle \epsilon^2 = \frac{1}{4}k^2\left( k^2\!-\!4a^2 \right) $} \\
		\cline{1-1}\cline{4-5}
		(iv) & & & $ a\cot(ax) $ & $\displaystyle \vphantom{\Bigg|} \mathrm{e}^{\mathrm{i}kx}\left( a\cot(ax)-\frac{\mathrm{i}k}{2} \right) $ & & \\
		\hline
		(v) & \multirow{3}*[-1.6em]{0} & \multirow{2}*[-0.9em]{$\displaystyle-\frac{b^2}{2}<0$} & $\displaystyle \frac{b}{\cos(bx)} $ & $\displaystyle \vphantom{\Bigg|} \mathrm{e}^{\mathrm{i}kx}\left( b\tan(bx)+\mathrm{i}\frac{k^2+b^2}{2k} \right) $ & \multirow{2}*[-0.9em]{$\displaystyle -\frac{\mathrm{i}(k^2-b^2)}{2\epsilon k}S' $} & \multirow{2}*[-0.9em]{$\displaystyle \epsilon^2 = \frac{1}{4}(k^2-b^2)^2 $} \\
		\cline{1-1}\cline{4-5}
		(vi) & & & $\displaystyle \frac{b}{\sin(bx)} $ & $\displaystyle \vphantom{\Bigg|} \mathrm{e}^{\mathrm{i}kx}\left( b\cot(bx)-\mathrm{i}\frac{k^2+b^2}{2k} \right) $ & & \\
		\cline{1-1}\cline{3-7}
		(vii) & & $\displaystyle\frac{b^2}{2}>0$ & $\displaystyle \frac{b}{\sinh(bx)} $ & $\displaystyle \vphantom{\Bigg|} \mathrm{e}^{\mathrm{i}kx}\left( b\coth(bx)-\mathrm{i}\frac{k^2-b^2}{2k} \right) $ & $\displaystyle -\frac{\mathrm{i}(k^2+b^2)}{2\epsilon k}S' $ & $\displaystyle \epsilon^2 = \frac{1}{4}(k^2+b^2)^2 $ \\
		\hline
		(viii) & 0 & $ 0 $ & $\displaystyle \frac{1}{x} $ & $\displaystyle \vphantom{\Bigg|} \mathrm{e}^{\mathrm{i}kx}\left( \frac{1}{x}-\frac{\mathrm{i}k}{2} \right) $ & $\displaystyle -\frac{\mathrm{i}k}{2\epsilon}S' $ & $\displaystyle \epsilon^2 = \frac{1}{4}k^4 $ \\
		\hline
		\end{tabular}
		}
		\end{center}
	\end{table*}
\section{Applications}\label{secapply}
	In this section, we treat two examples, that is, a potential step $ U(x)=U_0\theta(x) $ and a rectangular barrier $ U(x) = U_0\theta(a-|x|) $ \ ($ U_0,\,a\!>\!0 $), and investigate the transmission and reflection properties of Bogoliubov excitations.
\subsection{Rectangular Barrier}\label{subsec:rec}
\begin{figure}[tb]
		\begin{center}
		\includegraphics{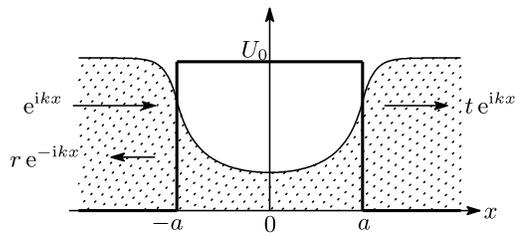}
		\caption{\label{fig:recset}Problem of tunneling of Bogoliubov excitations across a rectangular barrier. Shaded area represents the condensate wavefunction $ \Psi(x) $.}
		\end{center}
	\end{figure}

\begin{figure}[tb]
		\begin{center}
		\begin{tabular}{r}
		(a) \\[-1em]
		\includegraphics[scale=0.9]{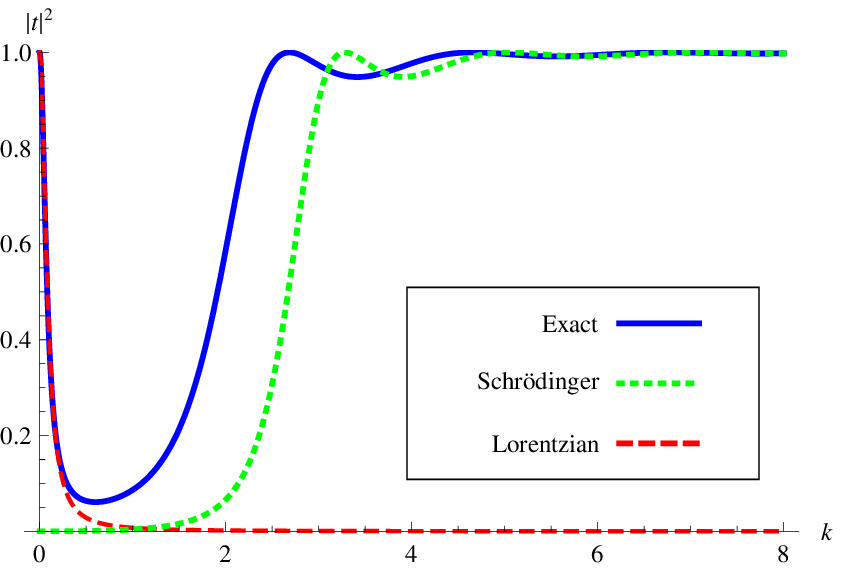}\\[0.75em]
		\end{tabular}
		\\
		\begin{tabular}{rr}
		(b) \ &(c) \ \\[-1.5em]
		\includegraphics[scale=0.8]{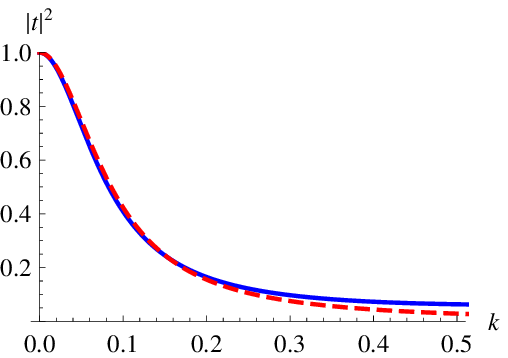} & \includegraphics[scale=0.8]{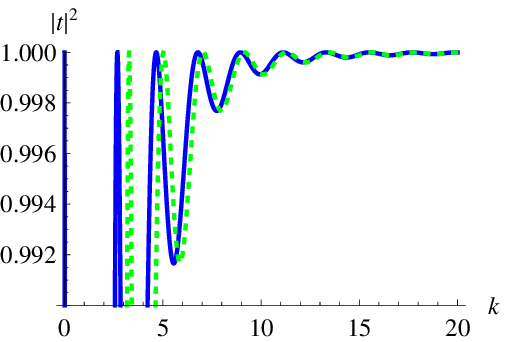}
		\end{tabular}
		\caption{\label{fig:rectransm}(Color online) (a): Transmission coefficients $ |t|^2 $ for a rectangular barrier with $ a=0.71,\,U_0=2.96 \ (\leftrightarrow b=0.2,\,\alpha=10)$. Horizontal axis represents not energy $ \epsilon $ but wavenumber $ k $. ``Schr\"odinger'' represents the solution for ordinary Schr\"odinger equation with the same potential barrier. ``Lorentzian'' curve is represented by Eq.~(\ref{eq:lorentz}).  (b): Close-up of low-energy region. (c): Close-up of high-energy region.}
		\end{center}
	\end{figure}

\begin{figure}[tb]
		\begin{center}
		\begin{tabular}{c}
		\includegraphics[scale=0.9]{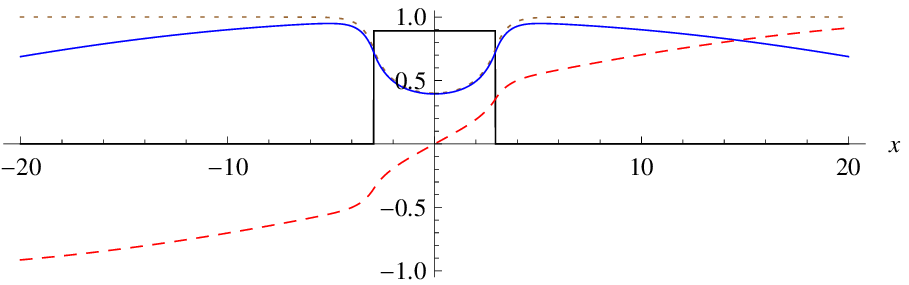} \\
		\includegraphics[scale=0.9]{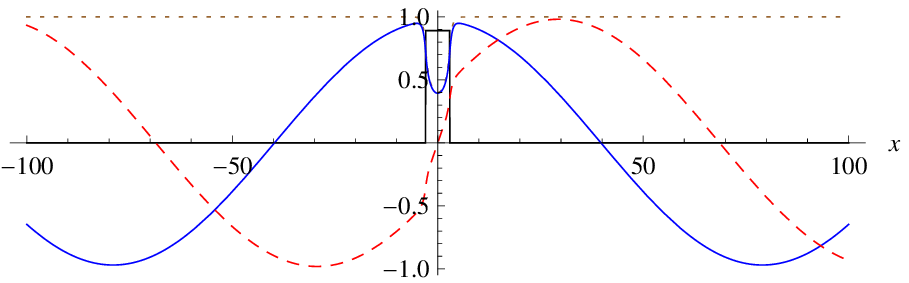} \\
		\includegraphics[scale=1]{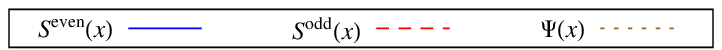}
		\end{tabular}
		\caption{\label{fig:recS}(Color online)  $ S^{\text{even}}(x) $ and $ S^{\text{odd}}(x) $ with $ a=2.9,\,U_0=0.89 \ (\leftrightarrow b=0.4,\,\alpha=0.8),\, \text{and } \epsilon=0.04 $. Upper and lower figures show the same plot with different ranges. Near the barrier, the even solution is quite similar to the condensate wavefunction. Far from the barrier, they behave as simple sine waves.}
		\end{center}
	\end{figure}

	As the first example, we consider the tunneling properties of Bogoliubov excitations across a rectangular barrier $ U(x) = U_0\theta(a-|x|) $. See Fig. \ref{fig:recset}. This problem has been first considered by Kagan \textit{et~al.}\cite{kagan}. However, Bogoliubov equations have been solved only numerically.\\
	\indent Imposing the boundary condition $ \Psi(x\rightarrow\pm\infty)=1 $, one obtains $ \mu=1 $. (This normalization is always possible for any condensate density $ n_0 $ at infinity, by taking $ \xi=\hbar/\sqrt{mgn_0} $ in Eq. (\ref{eq:xi}).) The solution of GP equation without node and current is then given by
	\begin{align}
		\Psi(x) &= \begin{cases} \Psi^{\text{in}}(x) &(|x|\le a) \\ \Psi^{\text{out}}(x) &(|x|\ge a) \end{cases}; \\
		\Psi^{\text{in}}(x) &= \frac{b}{\operatorname{cn}(\alpha bx|m)}, \label{eq:recin} \\
		\Psi^{\text{out}}(x) &= \tanh\!\left( |x|-a+\tanh^{-1}\gamma \right)
	\end{align}
	with
	\begin{align}
		m=1-\alpha^{-2},\quad \gamma = \!\sqrt{\frac{1+b^4(\alpha^2-1)}{2+b^2(\alpha^2-2)}}.
	\end{align}
	Here the height and width of the barrier are parametrized by $ 0< b \le 1 $ and $ 0<\alpha<+\infty $ as
	\begin{align}
		U_0 &= 1-b^2+\frac{b^2\alpha^2}{2}, \\
		a &= \frac{1}{b\alpha}\operatorname{cn}^{-1}\biggl(\frac{b}{\gamma}\,\bigg|\,m\biggr).
	\end{align}
	Henceforth, we mainly use $ b $ and $ \alpha $ for parametrization of the system rather than $ a $ and $ U_0 $, because of calculational convenience. 
	 The case $ \alpha=1 (\leftrightarrow m=0)$ is interesting since it can be solved elementarily by using (v) of Table \ref{table:elem}. However, we do not concentrate on this particular case only.\\
	\indent Next, let us consider the Bogoliubov equations. The solution set outside the barrier can be written by means of (i) of Table \ref{table:elem}. They are
	\begin{align}
		S^{\text{out}}_j(x) &= \mathrm{e}^{\mathrm{i}k_jx}\Bigl( \Psi^{\text{out}}(x)-\frac{\mathrm{i}k_j}{2}\operatorname{sgn}x \Bigr), \\
		G^{\text{out}}_j(x) &= -\frac{\mathrm{i}k_j}{2\epsilon}[S^{\text{out}}_j(x)]'.
	\end{align}
	Here  $ k_j\,(j=1,2,3,4) $ are defined as
	\begin{align}
	\begin{split}
		k_1 &= \sqrt{2(\sqrt{1+\epsilon^2}-1)},\quad k_2=-k_1, \\
		k_3 &= \mathrm{i} \sqrt{2(\sqrt{1+\epsilon^2}+1)},\quad \text{and } \ k_4=-k_3.
	\end{split}\label{eq:disper}
	\end{align}
	\indent The solution inside the barrier is no longer elementary, so we must solve Eq. (\ref{eq:1storderS}) for the elliptic function $ \Psi^{\text{in}}(x) $. Setting the proportionality constant in (\ref{eq:gds}) as $ K=2\epsilon/\kappa $, one obtains
	\begin{align}
	\begin{split}
		&S^{\text{in}}_j(x) = \sqrt{\Psi^{\text{in}}(x)^2+\frac{\kappa_j^2}{4}} \exp\Biggl[ \mathrm{i}\nu_j x+\\
		&\quad\, \mathrm{i}\Bigl( \frac{\kappa_j}{2}-\nu_j \Bigr)\frac{4b^2}{4b^2+\kappa_j^2}\frac{1}{b\alpha}\Pi\biggl(\!\frac{\kappa_j^2}{4b^2+\kappa_j^2};\operatorname{am}(\alpha bx|m)\Big|m\!\biggr)\!\Biggr]\!,
	\end{split}\\
		&G^{\text{in}}_j(x) = -\frac{\mathrm{i}\kappa_j}{2\epsilon}[S^{\text{in}}_j(x)]'
	\end{align}
	with
	\begin{align}
		\nu_j = \frac{4\epsilon^2}{\kappa_j^3}+\frac{b^2}{\kappa_j}(\alpha^2-2).
	\end{align}
	Here  $ \Pi(n;\varphi|m) $ and $ \operatorname{am}(u|m) $ are incomplete elliptic integral of the third kind and Jacobi amplitude, respectively. See Appendix \ref{appjacobi}. 
	 $ \kappa_j\,(j=1,2,3,4) $ are roots of the following equation:
	\begin{align}
		\epsilon^2 = \frac{1}{4}\left( \frac{4\epsilon^2}{\kappa^2}+b^2(\alpha^2-2) \right)^2+b^4(\alpha^2-1).
	\end{align}
	More explicitly, we define each $ \kappa_j $ as follows:
	\begin{align}
	\begin{split}
		\kappa_1 &= \sqrt{\frac{4\epsilon^2}{b^2(2-\alpha^2)+2\sqrt{\epsilon^2+b^4(1-\alpha^2)}}},\quad \kappa_2 = -\kappa_1, \\
		\kappa_3 &= \sqrt{\frac{4\epsilon^2}{b^2(2-\alpha^2)-2\sqrt{\epsilon^2+b^4(1-\alpha^2)}}},\ \text{and } \kappa_4 = -\kappa_3.
	\end{split}
	\end{align}
	\indent Since the sets of the solutions both inside and outside the barrier have been prepared, we can now construct the solution valid for all regions by joining them. Let us recall that our goal is to obtain the transmission amplitude, i.e., to construct the solution which has the following asymptotic form:
	\begin{align}
		\begin{pmatrix}S \\ G \end{pmatrix} = \left\{ \begin{aligned}&  \begin{pmatrix} 1 \\ k_1^2/(2\epsilon) \end{pmatrix}(\mathrm{e}^{\mathrm{i}k_1x}+ r\,\mathrm{e}^{-\mathrm{i}k_1x} ) & (x\rightarrow-\infty) \\ & \begin{pmatrix} 1 \\ k_1^2/(2\epsilon) \end{pmatrix}t\,\mathrm{e}^{\mathrm{i}k_1x} & (x\rightarrow+\infty) \end{aligned} \right. . \label{eq:recasym}
	\end{align}
	This asymptotic form defines the transmission amplitude $ t $ and reflection amplitude $ r $. These quantities are well-defined, as a result of the constancy of Wronskian:
	\begin{align}
		W=u^*\frac{\mathrm{d}u}{\mathrm{d}x}\!-\!u\frac{\mathrm{d}u^*}{\mathrm{d}x}\!+\!v^*\frac{\mathrm{d}v}{\mathrm{d}x}\!-\!v\frac{\mathrm{d}v^*}{\mathrm{d}x}. \label{wronskian}
	\end{align}
	Calculating $ W $ from Eq. (\ref{eq:recasym}), one obtains $ |t|^2+|r|^2=1 $. Physically, the constancy of Wronskian corresponds to the conservation of excitation energy \cite{kagan}.\\
	\indent Instead of constructing the solution with the form (\ref{eq:recasym}) directly, we first construct the even and odd solutions free from exponential divergence, because the calculation becomes a little easier. Even and odd solutions, $ (S^{\text{even}}\!,G^{\text{even}})\, $ and $ \,(S^{\text{odd}}\!,G^{\text{odd}}) $, should have the following form:
	\begin{align}
	F^{\text{even}} &=
		\begin{cases}
			c_2F^{\text{out}}_1+c_1F^{\text{out}}_2+c_3F^{\text{out}}_4 & (x<-a) \\
			c_4(F^{\text{in}}_1+F^{\text{in}}_2)+c_5(F^{\text{in}}_3+F^{\text{in}}_4) & (|x|\le a) \\
			c_1F^{\text{out}}_1+c_2F^{\text{out}}_2+c_3F^{\text{out}}_3 & (x>a)
		\end{cases},\\
	F^{\text{odd}} &=
		\begin{cases}
			-d_2F^{\text{out}}_1-d_1F^{\text{out}}_2-d_3F^{\text{out}}_4 & (x<-a) \\
			d_4(F^{\text{in}}_1-F^{\text{in}}_2)+d_5(F^{\text{in}}_3-F^{\text{in}}_4) & (|x|\le a) \\
			d_1F^{\text{out}}_1+d_2F^{\text{out}}_2+d_3F^{\text{out}}_3 & (x>a)
		\end{cases}
	\end{align}
	with $ F\!=\!S \text{ and }G $. 
	From the continuity conditions at $ x=a $, the coefficients $ c_i $ and $ d_i $ must satisfy 
	\begin{align}
	\begin{split}
		&f^{\text{out}}_1(a)c_1+f^{\text{out}}_2(a)c_2+f^{\text{out}}_3(a)c_3 \\ 
		& \quad =(f^{\text{in}}_1(a)+f^{\text{in}}_2(a))c_4+(f^{\text{in}}_3(a)+f^{\text{in}}_4(a))c_5,
	\end{split} \label{eq:recc} \\
	\begin{split}
		&f^{\text{out}}_1(a)d_1+f^{\text{out}}_2(a)d_2+f^{\text{out}}_3(a)d_3 \\
		& \quad =(f^{\text{in}}_1(a)-f^{\text{in}}_2(a))d_4+(f^{\text{in}}_3(a)-f^{\text{in}}_4(a))d_5,
	\end{split} \label{eq:recd}
	\end{align}
	with  $ f=S,\,S',\,G,\, \text{and } G'$. The conditions at $ x=-a $ are automatically satisfied as a result of $ f $ being even or odd. The solutions of these linear equations can be conveniently written down by the method in Appendix \ref{app:lin}.\\
	\indent Once even and odd non-divergent solutions are constructed as the above form, the solution of the transmission-reflection problem can be easily derived; one has only to make the linear combination of the two so that the term proportional to $ \mathrm{e}^{-\mathrm{i}k_1x} $ does not exist at $ x\rightarrow+\infty $. It is
	\begin{align}
		\begin{pmatrix} S \\ G \end{pmatrix} = d_2 \begin{pmatrix} S^{\text{even}} \\ G^{\text{even}} \end{pmatrix}-c_2\begin{pmatrix} S^{\text{odd}} \\ G^{\text{odd}} \end{pmatrix}.
	\end{align}
	The transmission and reflection amplitudes defined from the asymptotic form (\ref{eq:recasym}) are then given, respectively, by
	\begin{align}
		t &= \frac{1}{2} \ \frac{2-\mathrm{i}k_1}{2+\mathrm{i}k_1}\left( \frac{c_1}{c_2}-\frac{d_1}{d_2} \right), \label{eq:rectrans} \\
		r &= \frac{1}{2} \ \frac{2-\mathrm{i}k_1}{2+\mathrm{i}k_1}\left( \frac{c_1}{c_2}+\frac{d_1}{d_2} \right).
	\end{align}
	Now, we can plot the transmission coefficient $ |t|^2 $. See Fig.~\ref{fig:rectransm}. It shows perfect transmission in the limit $ \epsilon\rightarrow0 $, i.e., anomalous tunneling\cite{kovrizhinmaksimov,kovrizhin,kagan} occurs.\\
	\indent In order to verify the exact solution of Fig.~\ref{fig:rectransm}, let us compare it with reliable established theories in both low-energy and high-energy regions. ``Lorentzian'' in Fig.~\ref{fig:rectransm} shows the following curve:
	\begin{align}
		|t|^2=\frac{1}{1+\gamma^2k_1^2} ,\ \gamma = \frac{1}{2}\int_{-\infty}^{\infty}\!\!\mathrm{d}x\left( \frac{1}{\Psi^2}-1 \right). \label{eq:lorentz}
	\end{align}
	The author and Kato\cite{takahashiunpub} have shown that this expression fits well the transmission coefficient of Bogoliubov excitations against a high barrier with an arbitrary shape in the low-energy region. Indeed, Figure~\ref{fig:rectransm}(b) shows that this fitting works very well. As the excitation energy $ \epsilon $ becomes high, the transmission coefficient of Bogoliubov excitation comes close to that of an ordinary particle which obeys Schr\"odinger equation, as shown in Fig.~\ref{fig:rectransm}(c). This is because high-energy Bogoliubov quasiparticle has almost the same properties as an ordinary particle.\\ 
	\indent Figure \ref{fig:recS} shows $ S(x) $ with very small excitation energy $ \epsilon $. For small $ \epsilon $, $ G(x) $ is very small compared to $ S(x) $, so it is not shown in the figure. It means that for sufficiently low energy, $ u(x)\simeq v(x)\simeq S(x)/2 $. The behavior of even solution near the barrier is quite similar to that of the condensate. Indeed, Bogoliubov equations with $ \epsilon=0 $ always have the solution $ (S,G)=(\Psi,0) $ \cite{fetter}, and with the use of this solution, perfect transmission has been shown for arbitrary shape of potential barrier\cite{KatoNishiwakiFujita}.
\subsection{Potential Step}
\begin{figure}[tb]
	\begin{tabular}{ll}
	(a)  $ U_0\!<\!\mu $ & \\
	\begin{minipage}[c][6em][c]{12.5em} \includegraphics[scale=0.8]{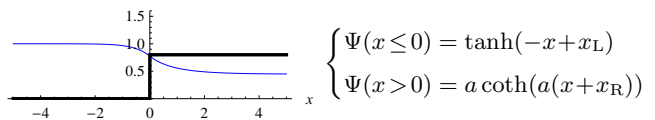}\end{minipage} & $ \displaystyle \begin{cases}\Psi(x\!\le\!0)=\tanh(-x\!+\!x_{\text{L}}) \\[1ex] \Psi(x\!>\!0)=a\coth(a(x\!+\!x_{\text{R}})) \end{cases} $ \\
	(b)  $ U_0\!=\!\mu $ & \\
	\begin{minipage}[c][6em][c]{12.5em} \includegraphics[scale=0.8]{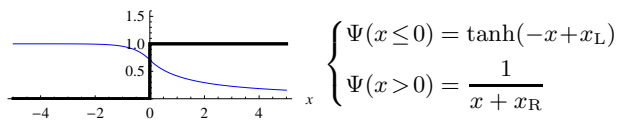}\end{minipage} & $ \displaystyle \begin{cases}\Psi(x\!\le\!0)=\tanh(-x\!+\!x_{\text{L}}) \\[1ex] \displaystyle \Psi(x\!>\!0)=\frac{1}{x+x_{\text{R}}} \end{cases} $ \\
	(c)  $ U_0\!>\!\mu $ & \\
	\begin{minipage}[c][6em][c]{12.5em} \includegraphics[scale=0.8]{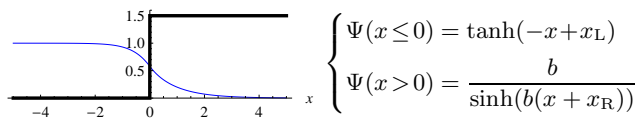}\end{minipage} & $ \displaystyle \begin{cases}\Psi(x\!\le\!0)=\tanh(-x\!+\!x_{\text{L}}) \\[1ex] \displaystyle \Psi(x\!>\!0)=\frac{b}{\sinh(b(x+x_{\text{R}}))} \end{cases} $
	\end{tabular}
		\caption{\label{fig:stepcond}(Color online) Condensate wavefunctions for the potential step $ U(x)=U_0\theta(x) $. The thick black lines represent the potential steps, and the thin blue lines represent the condensate wavefunctions. (a) $ U_0<\mu $. (b) $ U_0=\mu $. (c) $ U_0>\mu $. Note that $ \mu=1 $ in our normalization, and $ a=\sqrt{1-U_0},\, b=\sqrt{2(U_0-1)} $.}
	\end{figure}

	As the second example, let us consider the transmission and reflection properties of Bogoliubov excitations for a potential step $ U(x)=U_0\theta(x) $. We normalize the condensate density on the left side as $ 1 $, i.e., $ \Psi(x\!\rightarrow\!-\infty)=1 $, which determines the chemical potential as $ \mu=1 $.\\
	\indent Depending on the height of step $ U_0 $, the condensate can take the three forms, as shown in Fig.~\ref{fig:stepcond}. In the case (a), the height of the step is lower than the chemical potential, so the condensate on the right side has finite density. In the case (c), the condensate density vanishes in the limit $ x\rightarrow+\infty $ since the step is sufficiently high. The case (b) is the intermediate state of (a) and (c), and the condensate wavefunction shows power-law decay.
	 The constants $ x_{\text{R}} $ and $ x_{\text{L}} $ are determined from the continuity of $ \Psi(x) $ and $ \Psi'(x) $ at $ x\!=\!0 $. 
	They are
	\begin{align}
		x_{\text{L}} &= \begin{cases} \tanh^{-1}\bigl(\sqrt{(2-U_0)/2}\bigr) & (U_0 \le 1) \\  \tanh^{-1}\bigl(\sqrt{1/(2U_0)}\bigr) & (U_0>1) \end{cases}, \\
		x_{\text{R}} &= \begin{cases} \displaystyle \frac{1}{\sqrt{1-U_0}}\coth^{-1}\biggl(\sqrt{\frac{2-U_0}{2-2U_0}}\biggr) & (U_0 < 1) \\ \sqrt{2} & (U_0=1) \\ \displaystyle \frac{1}{\sqrt{2(U_0-1)}}\sinh^{-1}\bigl(2\sqrt{U_0(U_0-1)}\bigr) & (U_0>1) \end{cases}.
	\end{align}
	\indent Since the condensate wavefunction is expressed in terms of elementary functions, we can use Table \ref{table:elem} to solve the Bogoliubov equations. To avoid the complicated description, we prepare several notations before considering the transmission and reflection problems. As for the left side solutions, referring to (i) of Table \ref{table:elem}, we define the following:
	\begin{align}
		S^{\text{L}}_j(x) &= \mathrm{e}^{\mathrm{i}k^{\text{L}}_jx}\Bigl( \tanh(-x+x_{\text{L}})+\frac{\mathrm{i}k^{\text{L}}_j}{2} \Bigr), \\
		G^{\text{L}}_j(x) &= -\frac{\mathrm{i}k^{\text{L}}_j}{2\epsilon}[S^{\text{L}}_j(x)]'.
	\end{align}
	Here  $ k^{\text{L}}_j\,(j=1,2,3,4) $ are defined as
	\begin{align}
	\begin{split}
		k^{\text{L}}_1 &= \sqrt{2(\sqrt{1+\epsilon^2}-1)},\quad k^{\text{L}}_2=-k^{\text{L}}_1, \\
		k^{\text{L}}_3 &= \mathrm{i} \sqrt{2(\sqrt{1+\epsilon^2}+1)},\quad \text{and } \ k^{\text{L}}_4=-k^{\text{L}}_3.
	\end{split}
	\end{align}
	The notations for the right side $(x\!>\!0)$ are defined in the same way. In the case (a) of Fig. \ref{fig:stepcond}, referring to (ii) of Table \ref{table:elem},
	\begin{gather}
	\begin{align}
		S^{\text{R}}_j(x) &= \mathrm{e}^{\mathrm{i}k^{\text{R}}_jx}\Bigl( a \coth(a(x+x_{\text{R}}))-\frac{\mathrm{i}k^{\text{R}}_j}{2} \Bigr), \\
		G^{\text{R}}_j(x) &= -\frac{\mathrm{i}k^{\text{R}}_j}{2\epsilon}[S^{\text{R}}_j(x)]',
	\end{align}\\
		\begin{split}
		k^{\text{R}}_1 &= \sqrt{2(\sqrt{a^4+\epsilon^2}-a^2)},\quad k^{\text{R}}_2=-k^{\text{R}}_1, \\
		k^{\text{R}}_3 &= \mathrm{i} \sqrt{2(\sqrt{a^4+\epsilon^2}+a^2)},\quad \text{and } \ k^{\text{R}}_4=-k^{\text{R}}_3,
		\end{split}
	\end{gather}
	with $ a=\sqrt{1-U_0} $. In the case (b), referring to (viii) of Table \ref{table:elem},
	\begin{gather}
	\begin{align}
		S^{\text{R}}_j(x) &= \mathrm{e}^{\mathrm{i}k^{\text{R}}_jx}\Bigl( \frac{1}{x+\sqrt{2}}-\frac{\mathrm{i}k^{\text{R}}_j}{2} \Bigr), \\
		G^{\text{R}}_j(x) &= -\frac{\mathrm{i}k^{\text{R}}_j}{2\epsilon}[S^{\text{R}}_j(x)]',
	\end{align}\\
		\begin{split}
		k^{\text{R}}_1 &= \sqrt{2\epsilon},\quad k^{\text{R}}_2=-k^{\text{R}}_1, \\
		k^{\text{R}}_3 &= \mathrm{i} \sqrt{2\epsilon},\quad \text{and } \ k^{\text{R}}_4=-k^{\text{R}}_3.
		\end{split}
	\end{gather}
	In the case (c), referring to (vii) of Table \ref{table:elem},
	\begin{gather}
	\begin{align}
		S^{\text{R}}_j(x) &= \mathrm{e}^{\mathrm{i}k^{\text{R}}_jx}\Bigl( b\coth(b(x+x_{\text{R}}))-\mathrm{i}\frac{(k^{\text{R}}_j)^2-b^2}{2k^{\text{R}}_j} \Bigr), \\
		G^{\text{R}}_j(x) &= -\frac{\mathrm{i}\bigl((k^{\text{R}}_j)^2+b^2\bigr)}{2\epsilon k^{\text{R}}_j}[S^{\text{R}}_j(x)]',
	\end{align}\\
		\begin{split}
		k^{\text{R}}_1 &= \sqrt{2\epsilon-b^2},\quad k^{\text{R}}_2=-k^{\text{R}}_1, \\
		k^{\text{R}}_3 &= \mathrm{i} \sqrt{2\epsilon+b^2},\quad \text{and } \ k^{\text{R}}_4=-k^{\text{R}}_3,
		\end{split}
	\end{gather}
	with $ b=\sqrt{2(U_0-1)} $.\\
	\indent Now, let us construct the solution of transmission and reflection problem in the following form:
	\begin{align}
		\begin{pmatrix} S \\ G \end{pmatrix} = \left\{ \begin{aligned}& c_1 \begin{pmatrix} S^{\text{L}}_1 \\[0.25ex] G^{\text{L}}_1 \end{pmatrix}+ c_2 \begin{pmatrix} S^{\text{L}}_2 \\[0.25ex] G^{\text{L}}_2 \end{pmatrix}+ c_4 \begin{pmatrix} S^{\text{L}}_4 \\[0.25ex] G^{\text{L}}_4 \end{pmatrix} & (x<0) \\ & d_1 \begin{pmatrix} S^{\text{R}}_1 \\[0.25ex] G^{\text{R}}_1 \end{pmatrix}+d_3 \begin{pmatrix} S^{\text{R}}_3 \\[0.25ex] G^{\text{R}}_3 \end{pmatrix} & (x>0) \end{aligned} \right. \label{eq:steptunnel}
	\end{align}
	The coefficients are determined from the continuity of $ S,\,S',\,G,\,\text{and }G' $ at $ x\!=\!0 $. Since there are four equations for five variables, $ (c_1,c_2,c_4,d_1,d_3) $ is determined except for an overall factor. It can be explicitly written down by using the method in Appendix~\ref{app:lin}. The transmission coefficient is given by
	\begin{align}
		T = 1-R  = 1-\left| \frac{c_2}{c_1} \right|^2.
	\end{align}
	Here again, transmission and reflection coefficients are well-defined from the constancy of Wronskian(Eq.~(\ref{wronskian})). Note that $ |d_1/c_1|^2 $ is \textit{not} a transmission coefficient. \\
	\begin{figure}[tb]
		\begin{center}
		\includegraphics{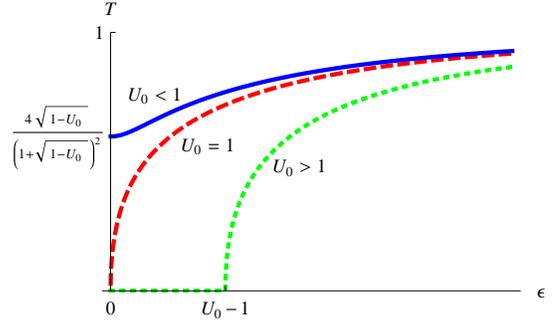}
		\caption{\label{fig:steptrans}(Color online) Transmission coefficients for step potential with step height $ U_0\lesseqqgtr1 $. Again note that $ \mu=1 $ in the present normalization.}
		\end{center}
	\end{figure}
	\begin{figure}[tb]
		\begin{center}
		\begin{tabular}{l}
		(a) $ U_0=0.9,\,\epsilon=0.2 $.  \\
		\includegraphics[scale=0.9]{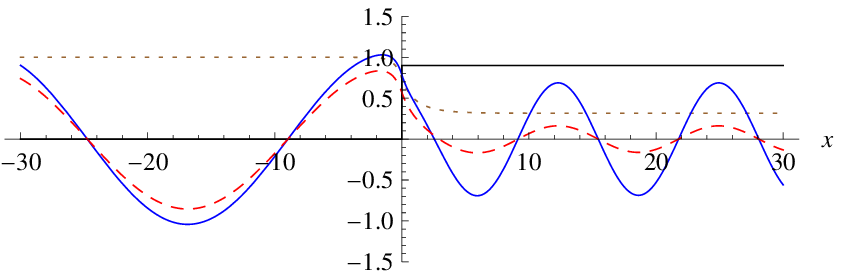} \\
		(b) $ U_0=1.3,\,\epsilon=0.25 $. \\
		\includegraphics[scale=0.9]{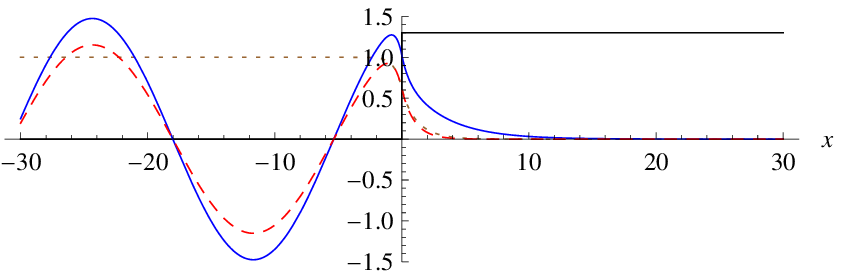} \\
		(c) $ U_0=1.3,\,\epsilon=0.4 $. \\
		\includegraphics[scale=0.9]{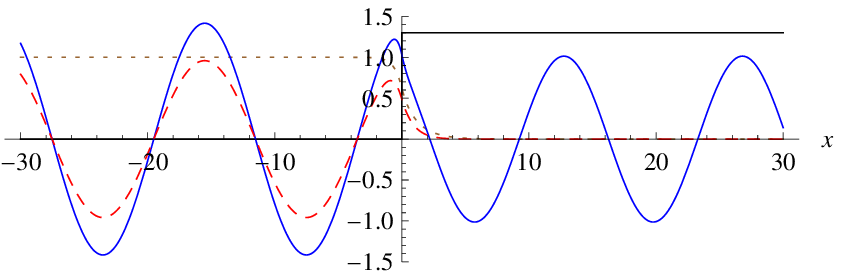} \\
		\includegraphics[scale=1]{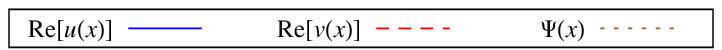}
		\end{tabular}
		\caption{\label{fig:stepuv}(Color online) Plot of $ \operatorname{Re}[u(x)] $ and $ \operatorname{Re}[v(x)] $ of the solutions (\ref{eq:steptunnel}) in various cases.  (a) $ U_0\!<\!1$.  (b) $ U_0\!>\!1 $ and $ \epsilon\!<\!U_0\!-\!1 $.  (c) $ U_0\!>\!1 $ and $ \epsilon\!>\!U_0\!-\!1 $. (Note that $ (u(x),v(x)) $ is multiplied by an overall phase factor so that the amplitudes of real and imaginary part become the same.)}
		\end{center}
	\end{figure}
	\indent The transmission coefficients are shown in Fig.~\ref{fig:steptrans}, and profiles of wavefunctions of excitations are shown in Fig.~\ref{fig:stepuv}. If the condensate remains finite on the right side, Bogoliubov excitations show partial transmission in the low energy limit. This situation is shown in Fig.~\ref{fig:stepuv}(a) and the case $ U_0\!<\!1 $ of Fig.~\ref{fig:steptrans}. Such a case is studied in detail in Refs.~\cite{WatabeKato,TsuchiyaOhashi2} in more general systems. The value $ T(\epsilon\!\rightarrow\!0)=4\sqrt{1-U_0}/(1+\sqrt{1-U_0})^2 $ is consistent with these earlier works. If the condensate decays on the right side because of high step, low-energy excitations show perfect reflection, as a free particle obeying Schr\"odinger equation does. It corresponds to Fig.~\ref{fig:stepuv}(b) and the case $ U_0\!>\!1 $ of Fig.~\ref{fig:steptrans}. The higher-energy excitations can transmit partially, but the component $ v(x) $ is suppressed on the right side, and only $ u(x) $ survives, as shown in Fig.~\ref{fig:stepuv}(c). This means that the quasiparticle is converted into an ordinary particle when it is ejected from the condensate to the vacuum. Such a phenomenon is called quantum evaporation and investigated in detail by several authors \cite{brownwyatt,dalfovo}. (However, the present model has no roton-like dispersion, unlike the superfluid helium~4.)
\section{Extension to the Complex-Valued Condensate Wavefunction}\label{sec:comp}
	In this section, we discuss the case where the condensate wavefunction takes a complex value. Like the real-valued case in Sec.~\ref{secreal}, Bogoliubov equations can be reduced to a first-order differential equation. Therefore, Bogoliubov equations with a piecewise constant potential can be always solved irrespective of the form of condensate wavefunction.
\subsection{Main Results}
	\indent Writing amplitude and phase of the condensate wavefunction as $ \Psi(x) = A(x) \exp[\mathrm{i}\Theta(x)] $, GP equation becomes
		\begin{gather}
		\hat{H}A=0,\ \hat{H}=-\frac{1}{2}\frac{\mathrm{d}^2 }{\mathrm{d} x^2}+U(x)-\mu+A^2+\frac{q^2}{2A^4}, \label{eq:GPamp0} \\
		\frac{\mathrm{d} \Theta}{\mathrm{d} x} = \frac{q}{A^2}, \label{eq:GPphase0}
	\end{gather}
	where the second equation is already integrated once, and the constant of integration $ q $ can be interpreted as the condensate current. We re-define $ S $ and  $ G $ as follows:
	\begin{align}
		S&=u\mathrm{e}^{-\mathrm{i}\Theta}\!+v\mathrm{e}^{\mathrm{i}\Theta}, \\
		G&=u\mathrm{e}^{-\mathrm{i}\Theta}\!-v\mathrm{e}^{\mathrm{i}\Theta}.
	\end{align}
	By using these quantities, Bogoliubov equations can be rewritten as
	\begin{align}
		\hat{H}S-\frac{\mathrm{i}q}{A}\frac{\mathrm{d}}{\mathrm{d}x}\!\left( \frac{G}{A} \right)\! = \epsilon G, \label{eq:bogoScomp} \\
		(\hat{H}+2A^2)G-\frac{\mathrm{i}q}{A}\frac{\mathrm{d}}{\mathrm{d}x}\!\left( \frac{S}{A} \right)\! = \epsilon S. \label{eq:bogoGcomp}
	\end{align}
	Henceforth, we consider the constant potential $ U(x)=U_0=\text{const} $. We can then integrate GP equation (\ref{eq:GPamp0}) once, and obtain
	\begin{align}
		(A')^2=A^4+2A^2(U_0-\mu)-\frac{q^2}{A^2}+C_{\text{GP}}. \label{eq:GPoncecomp}
	\end{align}
	The solution of this equation can be described in terms of elliptic functions \cite{seaman}.\\
	\indent With use of $ C_{\text{GP}} $ in Eq.~(\ref{eq:GPoncecomp}), we summarize the solution of Bogoliubov equations as follows:\\[0.5ex]
	\textit{\indent There exists a particular solution which satisfies the relation}
	\begin{align}
		\frac{\mathrm{d}S}{\mathrm{d}x} = \mathrm{i}\Bigl(K-\frac{\mathrm{d}\Theta}{\mathrm{d}x}  \Bigr)G = \mathrm{i}\Bigl(K-\frac{q}{A^2} \Bigr)G, \label{eq:gds2}
	\end{align}
	\textit{and $ S $ satisfies the following first-order differential equation:}
	\begin{align}
	\begin{split}
		& \Bigl[ \mathrm{i}\epsilon\left( KA^2+K\tau-q+(K^2-\tau)qA^{-2}-q^3A^{-6} \right) \\
		& \ \ +AA'\left( K^2+KqA^{-2}+q(K\tau-2q)A^{-4}-q^2\tau A^{-6} \right) \Bigr]S \\
		& \ = \Bigl[ \epsilon^2+3Kq+K^2A^2 \\
		& \qquad\qquad +q(2q+K\tau)A^{-2}+q^2\tau A^{-4}-\mathrm{i}q\epsilon A^{-3}A' \Bigr]S'
	\end{split}\label{eq:1storderScomp}
	\end{align}
	\textit{with $ \tau = K^2/2+U_0-\mu $. The constant $ K $ must satisfy the following quartic equation:}
	\begin{align}
		\epsilon^2 = \left( \frac{1}{2}K^2+U_0-\mu \right)^2-2Kq-C_{\text{GP}}. \label{eq:detkcomp}
	\end{align}
	\indent One can easily make sure that if one sets $ q=0 $ in Eqs.~(\ref{eq:GPoncecomp})-(\ref{eq:detkcomp}), they are reduced to Eqs.~(\ref{eq:GPonce})-(\ref{eq:detk}). Since Eq.~(\ref{eq:1storderScomp}) is first-order, it can be solved  by the method of separation of variables. Thus, it is proved that Bogoliubov equation with piecewise constant potential is always solvable.\\
	\indent From a practical viewpoint, however, this solution seems to be less useful compared to real-valued case, because the solution of Eq.~(\ref{eq:1storderScomp}) is difficult to express in terms of already known functions. This is in contrast to the real-valued case, where the solution can be expressed explicitly in terms of incomplete elliptic integral of the third kind. \\
	\indent Before closing this subsection, we sketch the proof of the above results briefly. One can show by a direct calculation that $ S'+\mathrm{i}\Theta'G $ and $ G $ obey the same fourth order differential equation, even though an operator identity corresponding to Eq.~(\ref{eq:opeiden}) no longer exists in the complex-valued case. Therefore the condition (\ref{eq:gds2}) follows. Eliminating $ G $ and high-order derivatives of $ S $ from Eqs.~(\ref{eq:gds2}), (\ref{eq:bogoScomp}), and (\ref{eq:bogoGcomp}), one obtains Eq.~(\ref{eq:1storderScomp}). Using Eqs.~(\ref{eq:bogoScomp}), (\ref{eq:GPoncecomp}), (\ref{eq:1storderScomp}), and derivative of Eq.~(\ref{eq:1storderScomp}), one obtains Eq.~(\ref{eq:detkcomp}).
\subsection{Check by the solution for gray soliton state}
	\indent Though the general solution of Eq.~(\ref{eq:1storderScomp}) is difficult to express, we can observe its validity by substitution of one particular example. Let us consider the gray soliton state. (We note that in soliton theory, the term ``gray soliton'' is rarely used, and gray soliton is also called as dark soliton.) The condensate wavefunction is given by
	\begin{gather}
		\Psi(x) = \mathrm{e}^{\mathrm{i}(q/\rho)x}\biggl[\frac{q}{\rho}+\mathrm{i} \sqrt{\rho-\frac{q^2}{\rho^2}}\tanh\Bigl(\sqrt{\rho-\frac{q^2}{\rho^2}}x\Bigr) \biggr], \\
		\therefore \ A(x)^2 = \frac{q^2}{\rho^2}+\left( \rho-\frac{q^2}{\rho^2} \right)\tanh^2\Bigl(\sqrt{\rho-\frac{q^2}{\rho^2}}x\Bigr).
	\end{gather}
	Here $ |q|^2\le \rho^3 $ holds. $ q $ and $ \rho $ represent the supercurrent and the density far from soliton, respectively. The solution of Bogoliubov equations for gray soliton is given in several papers\cite{chen,BilasPavloff,danshitayokoshikurihara}. If it is rewritten in terms of $ S $ and $ G $ instead of $ u $ and $ v $, one obtains
	\begin{align}
	\begin{split}
		S&=\mathrm{e}^{\mathrm{i}kx}\biggl[ \left( 1-\frac{qk}{2\epsilon\rho} \right)A+\frac{qk}{2\epsilon\rho}\left( \rho+\frac{k^2}{2} \right)A^{-1} \\
		& \qquad\qquad\qquad\qquad\quad -\frac{\mathrm{i}k}{2}\left( 1-\frac{qk}{\epsilon\rho} \right)\frac{A'}{\rho-A^2} \biggr],
	\end{split}\label{eq:grayS}
	\\
	\begin{split}
		G &= \mathrm{e}^{\mathrm{i}kx}\biggl[ \frac{k^2}{2\epsilon}A+\frac{qk}{2\epsilon\rho}\left( \epsilon-\frac{qk}{\rho} \right)A^{-1}\\
		&\qquad\qquad\qquad -\frac{\mathrm{i}k}{2\epsilon}\left( \rho+\frac{k^2}{2}-A^2 \right)\frac{A'}{\rho-A^2} \biggr],
	\end{split}\label{eq:grayG}
	\end{align}
	where $ k $ satisfies
	\begin{align}
		\epsilon = \frac{qk}{\rho}+\frac{1}{2}\sqrt{k^2(k^2+4\rho)}.
	\end{align}
	From expressions (\ref{eq:grayS}) and (\ref{eq:grayG}), one can show
	\begin{align}
		S' = \mathrm{i}\left( \frac{2\epsilon}{k}-\frac{q}{\rho}-\frac{q}{A^2} \right)G.
	\end{align}
	Furthermore, one can make sure by direct substitution that Eq.~(\ref{eq:1storderScomp}) with $ K=(2\epsilon/k)-(q/\rho) $ has the solution of the form (\ref{eq:grayS}).
\section{Discussion and Future Works}\label{secout}
\subsection{Relation to soliton theory and extension to spinor BEC}
	When we solve the Bogoliubov equations, Equations~(\ref{eq:gds}) and (\ref{eq:gds2}) are the most important ansatz. We have derived these conditions by showing that $ G $ and $ S' $ (in the complex-valued case, $ G $ and $ S'+\mathrm{i}\Theta'G $ ) obey the same differential equation. A more sophisticated understanding of these ansatz comes from Lax pair of soliton theory. In Ref.~\cite{chen}, it is shown that the solution of Bogoliubov equations can be represented by the square of the solution  of linear problem of Lax pair, called as ``squared Jost function''. Using this fact and one of linear equations (Eqs. (7) and (9) of their paper), one can derive the equation which is equivalent to our ansatz, namely, Eqs.~(\ref{eq:gds}) and (\ref{eq:gds2}).\\
	\indent The method of squared Jost function suggests the possibility of the extension to spinor BEC. In Refs.~\cite{IedaMiyakawaWadati,IedaMiyakawaWadati2,UchiyamaIedaWadati}, it is shown that spin-1 Gross-Pitaevskii equation becomes integrable when the strengths of interaction satisfy a certain condition. The integrability of this system originates from that of matrix nonlinear Schr\"odinger equation\cite{TsuchidaWadati,IedaUchiyamaWadati}. We conjecture that Bogoliubov equations of this integrable spin-1 BEC system can be also solved by means of squared Jost function, as well as scalar BEC. Investigation of this conjecture and application to the tunneling problem of collective modes are future works.
\subsection{Instability of soliton train states under the perturbative potential}
	In Sec.~\ref{secapply}, we have mainly applied the exact solutions to the transmission and reflection problems of collective excitations. Another important application of our solutions is a stability study of soliton train states\cite{CarrClark,seaman}. Stationary GP equation in a uniform and infinite system has the solution of the form $ \Psi(x) \propto \operatorname{sn}(ax|m) $. Under this snoidal condensate wavefunction, the solutions of Bogoliubov equations can be obtained by using Eq.~(\ref{eq:1storderS}) and Appendix~\ref{appjacobi}. From the explicit expressions of exact solutions, one can derive the following facts: (a)For positive energy $ 0<\epsilon<\epsilon_{\text{c}} $, where $ \epsilon_{\text{c}} $ is a certain positive value determined from chemical potential, two of the four linearly-independent solutions have ``positive norm'', i.e., $ |u(x)|^2-|v(x)|^2>0 $, and the other two have ``negative norm'', i.e., $ |u(x)|^2-|v(x)|^2<0 $. (Even though the solution in an infinite system cannot be normalized, the sign of $ |u(x)|^2-|v(x)|^2 $ is meaningful.) (b)For $ \epsilon>\epsilon_{\text{c}} $, two of the four solutions have ``positive norm'', and the other two are unphysical solutions which diverge exponentially. Since Bogoliubov equations always have the solution pair $(\epsilon, u,v)$ and $ (-\epsilon,v^*,u^*) $, the above (a) indicates that the soliton train state possesses Landau instability\cite{landau}. It should be noted that $ \epsilon_{\text{c}}$ goes to $0$ in the limit of  $ \operatorname{sn}(ax|m)\rightarrow \tanh(ax) $, i.e., one dark soliton limit. Therefore, one dark soliton state has no Landau instability. (This is clear from the explicit expression in (i) of Table~\ref{table:elem}.) Thus, the existence of negative energy \textit{and} positive norm eigenstate is a property specific to soliton train states. We further note that positive norm state and negative norm state degenerates.  As shown in Ref.~\cite{NakamuraMineOkumuraYamanaka}, this degeneracy is a necessary (not sufficient) condition for emergence of complex eigenvalue. Therefore, when perturbative potential becomes sufficiently strong, one can expect that the dynamical instability would appear, as demonstrated in a double well system\cite{IchiharaDanshitaNikuni}. This rough discussion on infinite system would be modified in a finite system, because eigenstates are discretized. In order to obtain more precise results, a more quantitative study for a finite system with perturbative potential is needed.
\subsection{Supercurrent state through a potential step}
	In the studies on physical origin of anomalous tunneling\cite{kovrizhinmaksimov,kovrizhin,kagan}, it was important to reduce the physical property of low-energy excitations to that of the condensate wavefunction. Indeed, the perfect transmission has been proved for generic potential by using the coincidence between the condensate wavefunction and the wavefunctions of excitations in the low-energy limit\cite{KatoNishiwakiFujita}. Furthermore, the similarity between the superfluidity of the condensate and the perfect transmission of low-energy excitations has been proposed\cite{OhashiTsuchiya}. These works suggest that the perfect transmission of excitations can be regarded as an extended concept of the superfluidity of the condensate.\\
	\indent In the presence of potential step, as shown in Refs.~\cite{WatabeKato,TsuchiyaOhashi2} and Sec.~\ref{secapply} of the present paper, the excitations show partial transmission at zero-energy. (See Fig.~\ref{fig:steptrans} again.) Recently, the present author has found that there exists a supercurrent state through a potential step without reflection, whose density profile is quite similar to Fig.~\ref{fig:stepcond}(a). We conjecture that Bogoliubov excitations show partial transmission in the low-energy limit in this supercurrent state, as the case without supercurrent does. If so, it gives a counterexample to the above earlier work\cite{OhashiTsuchiya} which relates the superfluidity of the condensate and the perfect transmission of the excitation. That is to say, the existence of the supercurrent state of the condensate does not necessarily mean the perfect transmission of the excitation. Thus, this problem may give a chance to reconsider the physical interpretation of anomalous tunneling effect. We will report on this issue elsewhere.
\section{Conclusion}\label{seccon}
	In this paper, we have shown that one-dimensional Bogoliubov equations with piecewise constant potentials can be reduced to a first-order linear differential equation, and therefore, it can be always solved by the method of separation of variables. Particularly, when the condensate wavefunction is a real-valued function, the solution of Bogoliubov equations can be expressed explicitly in terms of the elliptic integral. Using these solutions, we have solved transmission and reflection problems of excitations for a rectangular barrier and a potential step. Our results provide new exact examples of anomalous tunneling effect and quantum evaporation.
\begin{acknowledgments}
	The author would like to thank Y.~Kato, S. Watabe and Y. Nagai for helpful discussions. This research was partially supported by the Ministry of Education, Science, Sports and Culture, Grant-in-Aid for Scientific Research on Priority Areas, 20029007, and also partially supported by Japan Society of Promotion of Science, Grant-in-Aid for Scientific Research (C), 21540352.
\end{acknowledgments}
\appendix
\section{Formula for Incomplete Elliptic Integral of the third kind}\label{appjacobi}
	We only show the formula necessary in this paper. More information on elliptic integrals and elliptic functions is available, e.g., in Refs.~\cite{aands,mathfunc}.\\
	\indent Jacobi amplitude is defined as the inverse function of incomplete elliptic integral of the first kind:
	\begin{align}
		\operatorname{am}^{-1}(\varphi|m) &= \int_0^\varphi\!\!\frac{\mathrm{d}\theta}{\sqrt{1-m\sin^2\theta}}.
	\end{align}
	Jacobi elliptic functions are then defined as
	\begin{align}
		\operatorname{sn}(u|m)&=\sin(\operatorname{am}(u|m)), \\
		\operatorname{cn}(u|m)&=\cos(\operatorname{am}(u|m)), \\
		\operatorname{dn}(u|m)&=\frac{\partial}{\partial u}\operatorname{am}(u|m).
	\end{align}
	Incomplete elliptic integral of the third kind is defined as
	\begin{align}
		\Pi(n;\varphi|m)= \int_0^\varphi\!\!\frac{\mathrm{d}\theta}{(1-n\sin^2\theta)\sqrt{1-m\sin^2\theta}}.
	\end{align}
	Setting $ \varphi=\operatorname{am}(u|m) $ and $ \theta=\operatorname{am}(z|m) $, one obtains
	\begin{align}
		\Pi(n;\operatorname{am}(u|m)|m)= \int_0^u\!\!\frac{\mathrm{d}z}{1-n\operatorname{sn}^2(z|m)}. \label{eq:app3rd}
	\end{align}
	Since any squares of Jacobi elliptic functions are related to $ \operatorname{sn}^2 $, one can always solve Eq. (\ref{eq:1storderS}) by means of this formula.
\section{Solution of $ n-1 $ Homogeneous Linear Equations with  $ n $ Unknowns} \label{app:lin}
	Consider four homogeneous linear equations with five unknowns:
	\begin{align}
		\begin{pmatrix}
		a_{11} & a_{12} & a_{13} & a_{14} & a_{15} \\
		a_{21} & a_{22} & a_{23} & a_{24} & a_{25} \\
		a_{31} & a_{32} & a_{33} & a_{34} & a_{35} \\
		a_{41} & a_{42} & a_{43} & a_{44} & a_{45}
		\end{pmatrix}
		\begin{pmatrix}
		x_1 \\ x_2 \\ x_3 \\ x_4 \\ x_5
		\end{pmatrix}
		=\boldsymbol{0} \label{eqapp1}
	\end{align}
	Note that the coefficient matrix is \textit{not} square. Obviously, except for an overall factor, $(x_1,x_2,x_3,x_4,x_5)$ is determined uniquely. The solution is given by
	\begin{align}
		x_j \propto \Delta_j\,,
	\end{align}
	where $ \Delta_j $ is defined as
	\begin{align}
		\Delta_j = 
		\begin{vmatrix}
			a_{1,j+1} & a_{1,j+2} & a_{1,j+3} & a_{1,j+4} \\
			a_{2,j+1} & a_{2,j+2} & a_{2,j+3} & a_{2,j+4} \\
			a_{3,j+1} & a_{3,j+2} & a_{3,j+3} & a_{3,j+4} \\
			a_{4,j+1} & a_{4,j+2} & a_{4,j+3} & a_{4,j+4}
		\end{vmatrix}.
	\end{align}
	Here the matrix indices are considered by mod 5, i.e., $ j+n $ is replaced by $ j+n-5 $ if greater than 5. 
	It can be easily proved by cofactor expansion of the following trivially zero determinant:
	\begin{align}
		\begin{vmatrix}
		a_{i1} & a_{i2} & a_{i3} & a_{i4} & a_{i5} \\
		a_{11} & a_{12} & a_{13} & a_{14} & a_{15} \\
		a_{21} & a_{22} & a_{23} & a_{24} & a_{25} \\
		a_{31} & a_{32} & a_{33} & a_{34} & a_{35} \\
		a_{41} & a_{42} & a_{43} & a_{44} & a_{45}
		\end{vmatrix}
		=
		\sum_{j=1}^5 a_{ij}\Delta_j=0 \quad (i=1,2,3,4).
	\end{align}
	Thus, $ x_j=\Delta_j $ is the solution of Eq.~(\ref{eqapp1}). \\
	\indent When one of $ x_j $, say $ x_5 $, is fixed to a certain value, this formula is reduced to well-known Cramer's rule. Generalization of the above result to $ n-1 $ homogeneous linear equations with $ n $ unknowns is straightforward.\\
	\indent This formula is convenient to write down the solution of linear equations in a compact form. For example, $ c_1 $ and $ c_2 $ in Eq.~(\ref{eq:recc}) are given, respectively, by
	\begin{align}
		c_1 &=
 		\begin{vmatrix}
		S^{\text{out}}_2 & S^{\text{out}}_3& -S^{\text{in}}_1-S^{\text{in}}_2& -S^{\text{in}}_3-S^{\text{in}}_4 \\
		S^{\text{out}\,\prime}_2 & S^{\text{out}\,\prime}_3& -S^{\text{in}\,\prime}_1-S^{\text{in}\,\prime}_2& -S^{\text{in}\,\prime}_3-S^{\text{in}\,\prime}_4 \\
		G^{\text{out}}_2 & G^{\text{out}}_3& -G^{\text{in}}_1-G^{\text{in}}_2& -G^{\text{in}}_3-G^{\text{in}}_4 \\
		G^{\text{out}\,\prime}_2 & G^{\text{out}\,\prime}_3& -G^{\text{in}\,\prime}_1-G^{\text{in}\,\prime}_2& -G^{\text{in}\,\prime}_3-G^{\text{in}\,\prime}_4 \\
		\end{vmatrix},
		\\
		c_2 &=
		\begin{vmatrix}
		S^{\text{out}}_3& -S^{\text{in}}_1-S^{\text{in}}_2& -S^{\text{in}}_3-S^{\text{in}}_4 & S^{\text{out}}_1 \\
		S^{\text{out}\,\prime}_3& -S^{\text{in}\,\prime}_1-S^{\text{in}\,\prime}_2& -S^{\text{in}\,\prime}_3-S^{\text{in}\,\prime}_4 & S^{\text{out}\,\prime}_1 \\
		G^{\text{out}}_3& -G^{\text{in}}_1-G^{\text{in}}_2& -G^{\text{in}}_3-G^{\text{in}}_4 & G^{\text{out}}_1 \\
		G^{\text{out}\,\prime}_3& -G^{\text{in}\,\prime}_1-G^{\text{in}\,\prime}_2& -G^{\text{in}\,\prime}_3-G^{\text{in}\,\prime}_4 & G^{\text{out}\,\prime}_1 \\
		\end{vmatrix},
	\end{align}
	where the arguments of all matrix elements are $ a $, which are omitted to save space. $ d_1 $ and $ d_2 $ in Eq.~(\ref{eq:recd}) can be obtained in the same way. By using these expressions, the transmission coefficient (\ref{eq:rectrans}) can be written down compactly and explicitly. 
\bibliography{loverspj}

\end{document}